\documentclass[sigconf]{acmart}

\renewcommand\footnotetextcopyrightpermission[1]{} 
\usepackage{subcaption}

\newtheorem{myexample}{Example}

\newcommand{\ket}[1]{\ensuremath{\left|#1\right\rangle}}
\usepackage{qcircuit}
\usepackage{stmaryrd}

\usepackage{tikz}
\usetikzlibrary{arrows.meta}
\usetikzlibrary{shadows}
\usetikzlibrary{arrows,backgrounds,calc,chains,
	matrix,positioning,shapes,shapes.geometric,decorations,
	shapes.arrows, decorations.pathmorphing, decorations.pathreplacing, decorations.markings}

\tikzset{%
	font={\footnotesize},
	vertex/.style={draw,circle,inner sep=0pt,minimum width=0.5cm,minimum height=0.5cm},
	zeroterm/.style={below,inner sep=0pt,font=\tiny}
}

\let\OLDthebibliography\thebibliography
\renewcommand\thebibliography[1]{
	\OLDthebibliography{#1}
	\setlength{\parskip}{0pt}
	\setlength{\itemsep}{1.2pt plus 0.0ex}
}

\begin{document}
\date{}

\title{Compiling SU(4) Quantum Circuits to IBM QX Architectures}	

\author{
	Alwin Zulehner \hspace{3cm} Robert Wille \\
	{\normalsize Institute for Integrated Circuits, Johannes Kepler University Linz, Austria}\\
	{\normalsize alwin.zulehner@jku.at \hspace{3cm} robert.wille@jku.at}
}

\begin{abstract}
	The \emph{Noisy Intermediate-Scale Quantum}~(NISQ) technology
	is currently investigated by major players in the field to build the first practically useful quantum computer.
	\emph{IBM QX architectures} are the first ones which are already publicly available today. 
	However, in order to use them, the respective quantum circuits have to be \emph{compiled} for the respectively used target architecture. 
	While first approaches have been proposed for this purpose, they are infeasible for a certain set of \emph{SU(4)} quantum circuits 
	which have recently been introduced to benchmark corresponding compilers. 
	In this work, we analyze the bottlenecks of existing compilers and provide a dedicated method for compiling this kind of circuits to IBM QX architectures. Our experimental evaluation (using tools provided by IBM) shows that the proposed approach significantly outperforms IBM's own solution regarding fidelity of the compiled circuit as well as runtime. Moreover, the solution proposed in this work has been declared winner of the \emph{IBM QISKit Developer Challenge}.	
	An implementation of the proposed methodology is publicly available at \url{http://iic.jku.at/eda/research/ibm_qx_mapping}.
\end{abstract}

\maketitle
\thispagestyle{empty}

\section{Introduction}
\label{sec:intro}

Quantum computers offer a promising computation paradigm that allows to solve certain tasks significantly faster than conventional machines. Instead of bits, these devices operate on so-called \emph{qubits} that can not only be in one of the basis states $\ket{0}$ and $\ket{1}$, but also in an (almost) arbitrary superposition of both, i.e.~\mbox{$\ket{\phi} = \alpha\ket{0} + \beta\ket{1}$}. In combination with other quantum mechanical phenomena like entanglement and phase shifts, this allows to develop quantum circuits (i.e.~a sequence of operations that are applied to the qubits) that gain an exponential speedup compared to conventional machines for several practically relevant problems.

Currently, there is an ongoing ``race'' to build the first practically useful quantum computer between large companies like IBM, Intel, Rigetti, and Google~\cite{gomes2018quantumcomputing,hsu2018quantumcomputing,DBLP:conf/icrc/SeteZR16,courtland2017google}. 
They all develop devices that can be classified to the \emph{Noisy Intermediate-Scale Quantum} (NISQ~\cite{preskill2018quantum}) technology.
Although still limited by their number of available qubits and low fidelity, these devices provide the capability of running quantum algorithms for dedicated problems in domains such as quantum chemistry or physical simulation and they provide the first step towards \mbox{fault-tolerant} quantum computing. Among the different solutions currently developed by the companies mentioned above, IBM's approach (yielding so-called \emph{IBM QX architectures}) is the first one
which is already publicly available today (through a cloud access launched within their project \emph{IBM~Q}~\cite{ibmQ}). Because of this, we are focusing on this architecture in the following. 

However, in order to use IBM QX devices (or NISQ devices in general), the respective quantum circuits have to be compiled to the target architecture.
This includes a decomposition of the operations into elementary gates provided by the architecture, as well as a mapping procedure that maps the logical qubits of the circuit to the physical ones of the QX device. While for the decomposition step, several solutions exist (cf.~\cite{DBLP:journals/tcad/AmyMMR13,MWZ:2011,matsumoto2008representation,WSOD:2013}), especially the mapping step  constitutes a tough challenge, since further physical constraints have to be considered. 
In fact, 2-qubit gates can be applied to certain pairs of physical qubits only. Therefore, SWAP operations have to be inserted that exchange the state of two physical qubits and, by this, allow to ``move'' the logical qubits to positions where they can interact with each other. Since each additional operation further decreases the fidelity of the quantum circuit, their number shall be kept as small as possible. 

Accordingly, researchers investigated how to efficiently accomplish that---yielding a large body of solutions for minimizing the number of SWAP operations required for  satisfying the  physical constraints. 
But most of them (e.g.~the ones proposed in~\cite{DBLP:journals/tcad/WilleLD14,DBLP:conf/aspdac/ShafaeiSP14,DBLP:conf/aspdac/WilleKWRCD16,zulehner2017exact,bhattacharjee2017depth})
focus on so-called \emph{nearest neighbor constraints}, which are not sufficient to get executed on IBM QX architectures (or NISQ architectures in general for that purpose). Other ones (such as proposed in~\cite{fowler2004implementation,DBLP:journals/corr/VenturelliDRF17}) focus on specific quantum circuits only. In fact, to the best of our knowledge, only IBM's own solution~\cite{qiskit} (provided in the corresponding SDK) as well as the approaches recently proposed in~\cite{zulehner2017efficient,siraichi2018qubit} are capable of sufficiently compiling quantum circuits for IBM QX architectures thus far.

However, recently a set of quantum circuits (called \emph{SU(4) quantum circuits} in the following) has been introduced which turns out to constitute a worst case for these compiling methods---making them infeasible. This is a crucial issue since this kind of circuits has explicitly been advocated by IBM to benchmark compilers (e.g.~through a so-called QISKit Developer Challenge~\cite{ibmDeveloperChallenge}). Hence, for a class of circuits which is considered to be important by a major player in the development of quantum computers, no method exists for efficiently compiling them to IBM QX architectures.

In this paper, we address this problem by providing a \emph{dedicated} compiler for SU(4) quantum circuits for IBM QX architectures. 
To this end, we analyze the existing compilation approaches and determine their respective advantages and bottlenecks. Based on that evaluation, we present a compilation approach which explicitly takes the structure of SU(4) quantum circuits into consideration. 
Experimental evaluations clearly show that the proposed approach significantly outperforms IBM's current solution as well as the other recently provided compilers with respect to fidelity of the resulting circuits as well as regarding runtime. Moreover, the proposed approach has been declared winner of the IBM QISKit Developer Challenge. According to IBM, the proposed solution yields compiled circuits with at least 10\% better costs than the other submissions while generating them at least 6 times faster. 

The remainder of this work is structured as follows. In Section~\ref{sec:background}, we review IBM's QX architectures, the considered SU(4) quantum circuits, as well as the compilation problem itself. In Section~\ref{sec:motivation}, we review the existing state of the art discuss why existing solutions suffer in compiling SU(4) circuits---providing the motivation of this work. In Section~\ref{sec:proposed}, we present the dedicated solution in detail; followed by an experimental comparison to the state of the art in Section~\ref{sec:exp}. Section~\ref{sec:conclusions} concludes the paper.

\section{Background}
\label{sec:background}

In this section, we briefly discuss IBM's QX architectures as well as the considered quantum circuits and provide a more detailed description of the considered compilation task.

\subsection{IBM's QX Architectures}
\label{sec:architectures}

In 2017, IBM started the initiative \emph{IBM Q} in order to make quantum computers available to the broad audience via cloud access. Currently, their infrastructure contains two 5-qubit quantum devices located in Yorktown and Tenerife (also called \emph{IBM QX2} and \emph{IBM QX4}, respectively), as well as a 16-qubit device located in Rueschlikon (also called \emph{IBM QX5}), which are publicly available. Moreover, there exists a 20-qubit architecture located in Austin that is available for IBM's partners  and members of the IBM Q network. 

All these devices use superconducting qubits that are connected with coplanar waveguide bus resonators~\cite{qxbackends}. Quantum operations are conducted by applying microwave impulses to the qubits. By this, all these architectures have the same (or at least similar) physical constraints that have to be satisfied when running quantum algorithms (i.e.~\emph{quantum circuits}) on them.

In fact, IBM's QX architectures only support two types of quantum operations (i.e.~quantum gates): $U(\theta, \phi	,\lambda)=R_z(\phi)R_y(\theta)R_z(\lambda)$ is a single qubit gate, which is composed of two rotations around the \mbox{$z$-axis} and one rotation around the $y$-axis (i.e.~an Euler decomposition). Furthermore, a controlled NOT gate (i.e.~a \emph{CNOT}) can be applied to a pair of qubits. If the so-called control qubit (denoted as $\bullet$ in quantum circuits) is in basis state $\ket{1}$, the state of the target qubit (denoted as $\oplus$ in quantum circuits) is inverted. These two quantum gates provide a universal basis, i.e.~any quantum algorithm can be conducted by using $U$ and CNOT gates only.

However, besides the restriction regarding the available gates, there are further physical constraints given by the architecture.
In fact, CNOT gates can be
applied only to qubits that are connected by a bus resonator. Furthermore, only the qubit with lower frequency may serve as target while only the qubit with the higher frequency may serve as control (except for certain cases; cf.~\cite{qxbackends}). These restrictions are summarized in so-called \emph{coupling maps}. 

\begin{figure}
	\centering
	\begin{subfigure}[b]{0.49\linewidth}
		\centering
		\scalebox{0.85}{
			\begin{tikzpicture}[terminal/.style={draw,rectangle,inner sep=2pt}]
			\matrix[matrix of nodes,ampersand replacement=\&,every node/.style={vertex},column sep={0.75cm,between origins},row sep={.7cm,between origins}] (qmdd) {
				\node (n0) {$Q_0$}; \& \& \node (n1) {$Q_1$}; \\
				\& \node (n2) {$Q_2$}; \&  \\
				\node (n4) {$Q_4$}; \& \& \node (n3) {$Q_3$}; \\
			};
			\draw[-Implies,line width=0.5pt,double distance=2pt] (n0.east) -- (n1.west);
			\draw[-Implies,line width=0.5pt,double distance=2pt] (n0) -- (n2);
			\draw[-Implies,line width=0.5pt,double distance=2pt] (n1) -- (n2);
			\draw[-Implies,line width=0.5pt,double distance=2pt] (n3) -- (n2);
			\draw[-Implies,line width=0.5pt,double distance=2pt] (n3.west) -- (n4.east);
			\draw[-Implies,line width=0.5pt,double distance=2pt] (n4) -- (n2);
			\end{tikzpicture}}
		
		\vspace*{-2mm}
		\caption{\emph{QX2}}\label{fig:qx2arch}
	\end{subfigure}\hfill
	\begin{subfigure}[b]{0.49\linewidth}
		\centering
		\scalebox{0.85}{
			\begin{tikzpicture}[terminal/.style={draw,rectangle,inner sep=2pt}]
			\matrix[matrix of nodes,ampersand replacement=\&,every node/.style={vertex},column sep={0.75cm,between origins},row sep={0.7cm,between origins}] (qmdd) {
				\node (n0) {$Q_0$}; \& \& \node (n1) {$Q_1$}; \\
				\& \node (n2) {$Q_2$}; \&  \\
				\node (n4) {$Q_4$}; \& \& \node (n3) {$Q_3$}; \\
			};
			\draw[-Implies,line width=0.5pt,double distance=2pt] (n1.west) -- (n0.east);
			\draw[-Implies,line width=0.5pt,double distance=2pt] (n2) -- (n0);
			\draw[-Implies,line width=0.5pt,double distance=2pt] (n2) -- (n1);
			\draw[-Implies,line width=0.5pt,double distance=2pt] (n2) -- (n4);
			\draw[-Implies,line width=0.5pt,double distance=2pt] (n3.west) -- (n4.east);
			\draw[-Implies,line width=0.5pt,double distance=2pt] (n3) -- (n2);
			\end{tikzpicture}}
		
		\vspace*{-2mm}
		\caption{\emph{QX4}}\label{fig:qx4arch}
	\end{subfigure}

	\begin{subfigure}[b]{\linewidth}
		\centering
		\scalebox{0.85}{
			\begin{tikzpicture}[terminal/.style={draw,rectangle,inner sep=2pt}]
			\matrix[matrix of nodes,ampersand replacement=\&,every node/.style={vertex},column sep={1.25cm,between origins},row sep={1cm,between origins}] (qmdd) {
				\node (n1) {$Q_{1\phantom{0}}$}; \& \node (n2) {$Q_{2\phantom{0}}$}; \& \node (n3) {$Q_{3\phantom{0}}$}; \& \node (n4) {$Q_{4\phantom{0}}$}; \& \node (n5) {$Q_{5\phantom{0}}$}; \& \node (n6) {$Q_{6\phantom{0}}$}; \& \node (n7) {$Q_{7\phantom{0}}$}; \& \node (n8) {$Q_{8\phantom{0}}$}; \\
				\node (n0) {$Q_{0\phantom{0}}$}; \& \node (n15) {$Q_{15}$}; \& \node (n14) {$Q_{14}$}; \& \node (n13) {$Q_{13}$}; \& \node (n12) {$Q_{12}$}; \& \node (n11) {$Q_{11}$}; \& \node (n10) {$Q_{10}$}; \& \node (n9) {$Q_{9\phantom{0}}$}; \\
			};
			\draw[-Implies,line width=0.5pt,double distance=2pt] (n1.south)--(n0.north);
			\draw[-Implies,line width=0.5pt,double distance=2pt] (n1.east)--(n2.west);
			\draw[-Implies,line width=0.5pt,double distance=2pt] (n2.east)--(n3.west);		
			\draw[-Implies,line width=0.5pt,double distance=2pt] (n3.south)--(n14.north);
			\draw[-Implies,line width=0.5pt,double distance=2pt] (n3.east)--(n4.west);
			\draw[-Implies,line width=0.5pt,double distance=2pt] (n5.west)--(n4.east);
			\draw[-Implies,line width=0.5pt,double distance=2pt] (n6.west)--(n5.east);
			\draw[-Implies,line width=0.5pt,double distance=2pt] (n6.east)--(n7.west);
			\draw[-Implies,line width=0.5pt,double distance=2pt] (n6.south)--(n11.north);
			\draw[-Implies,line width=0.5pt,double distance=2pt] (n7.south)--(n10.north);
			\draw[-Implies,line width=0.5pt,double distance=2pt] (n8.west)--(n7.east);
			\draw[-Implies,line width=0.5pt,double distance=2pt] (n9.north)--(n8.south);
			\draw[-Implies,line width=0.5pt,double distance=2pt] (n9.west)--(n10.east);
			\draw[-Implies,line width=0.5pt,double distance=2pt] (n11.east)--(n10.west);
			\draw[-Implies,line width=0.5pt,double distance=2pt] (n12.east)--(n11.west);
			\draw[-Implies,line width=0.5pt,double distance=2pt] (n12.north)--(n5.south);
			\draw[-Implies,line width=0.5pt,double distance=2pt] (n12.west)--(n13.east);
			\draw[-Implies,line width=0.5pt,double distance=2pt] (n13.north)--(n4.south);
			\draw[-Implies,line width=0.5pt,double distance=2pt] (n13.west)--(n14.east);
			\draw[-Implies,line width=0.5pt,double distance=2pt] (n15.east)--(n14.west);
			\draw[-Implies,line width=0.5pt,double distance=2pt] (n15.west)--(n0.east);
			\draw[-Implies,line width=0.5pt,double distance=2pt] (n15.north)--(n2.south);
			\end{tikzpicture}
		}
		
		\vspace*{-2mm}
		\caption{\emph{QX5}}\label{fig:qx5arch}
	\end{subfigure}

		\vspace*{-3mm}
	\caption{Coupling map of the \emph{IBM QX} architectures~\cite{qxbackends}}
	\label{fig:coupling}
		\vspace*{-3mm}
\end{figure}
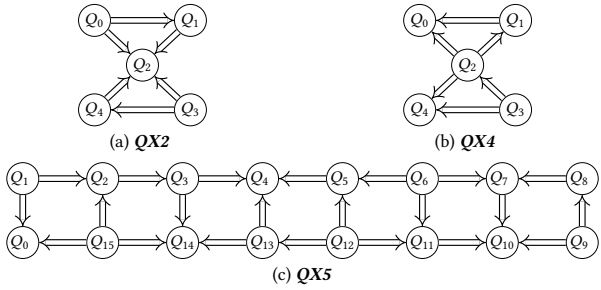

\begin{myexample}
Fig.~\ref{fig:coupling} shows the coupling maps for the \emph{IBM~QX2}, \emph{IBM~QX4}, and \emph{IBM~QX5} architectures. Here, qubits are visualized with vertices and an arrow pointing from qubit $Q_i$ to qubit $Q_j$ indicates that only CNOTs with control qubit $Q_i$ and target qubit $Q_j$ can be applied.
\end{myexample}

In the following, we denote the devices listed above, as well as (future) devices that employ the same type of constraints as \emph{IBM QX architectures}. Besides that, note that, since quantum computers are still in their infancy, applying a quantum gate fails with a certain probability (cf.~NISQ devices~\cite{preskill2018quantum}). According to data provided by IBM~\cite{ibmDevices}, CNOT operations approximately have  a fidelity that is 10 times smaller than for single qubit gates. Because of that, it is of uttermost importance to keep the number of CNOT gates in particular as small as possible.

\subsection{Considered Quantum Circuits}
\label{sec:circuits}

\begin{figure}
\begin{center}	
	\footnotesize
\mbox{\Qcircuit @C=.75em @R=0.5em {
\push{\rule{1em}{0em}} & \lstick{q_0} & \qw & \multigate{1}{SU(4)} & \qw &  \push{\rule[0mm]{.3em}{0em}\phantom{\equiv}\rule{.8em}{0em}} & \gate{U_1} & \ctrl{1} & \gate{U_3} & \targ & \gate{U_5} & \ctrl{1} & \gate{U_6} & \qw \\
\push{\rule{1em}{0em}} & \lstick{q_1} & \qw & \ghost{SU(4)} & \qw &  \push{\rule[-5.5mm]{.3em}{0em}\equiv\rule{.8em}{0em}}  & \gate{U_2} & \targ & \gate{U_4} & \ctrl{-1} & \qw & \targ & \gate{U_7}  &  \qw \\
	}}
\end{center}

		\vspace*{-3mm}
	\caption{KAK decomposition of an SU(4) gate}
	\label{fig:kak}
		\vspace*{-3mm}
\end{figure}
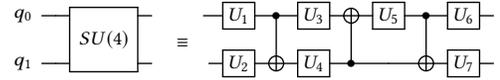

Quantum algorithms or quantum circuits are usually described using high-level quantum languages~\cite{abhari2012scaffold,DBLP:conf/pldi/GreenLRSV13}, quantum assembly languages (e.g.~OpenQASM 2.0 developed by IBM~\cite{cross2017open}), or circuit diagrams (such as those shown in Fig.~\ref{fig:kak}), where the qubits are visualized as circuit lines that are passed through quantum gates. These 
lines do not refer to an actual hardware connection (as in conventional logic), but rather define in which order (from left to right) the respective gates (i.e.~operations) are applied.

In this paper, we consider quantum circuits provided by IBM to benchmark the performance of respective compilers 
(e.g.~through a so-called QISKit Developer Challenge~\cite{ibmDeveloperChallenge}).
These circuits are products of random 2-qubit gates from SU(4) that are applied to random pairs of qubits and denoted as \emph{SU(4) quantum circuits} in the following.\footnote{SU(4) is the \emph{special unitary group with degree 4}, i.e.~the Lie group of $4\times 4$ unitary matrices with determinant 1. The functionality of any 2-qubit gate is described by an element from this group.} More precisely, in each layer of the circuit, the available qubits are grouped randomly into pairs of 2 qubits each (if their number is even). Then, to each of these pairs of qubits, a random 2-qubit gate from SU(4) is applied. Since these 2-qubit gates are not available in the gate set of the IBM QX architectures, \mbox{KAK-decomposition}~\cite{vatan2004optimal} is used to decompose each of these \mbox{2-qubit} gates into a sequence of three CNOTs and 7 single qubit gates. Eventually, these decomposed gates form the circuits for determining the performance of the compilers.

\begin{myexample}
	Fig.~\ref{fig:kak} shows the KAK decomposition of a random SU(4) gate. For simpler visualization, we neglect the parameters $\theta$, $\phi$, and $\lambda$ for the single qubit gates $U_i$ (which are usually different for each $U_i$). As can be seen, single qubit gates and CNOT gates are applied in an interleaved fashion.
\end{myexample}

\subsection{Considered Problem}
\label{sec:problem}

In this work, we consider how to efficiently compile 
the quantum circuits reviewed in the previous section to IBM QX architectures. In general, compilation is comprised of two steps. First, the operations occurring in the quantum circuits have to be decomposed into elementary operations that are available on the target hardware. In the literature, there exist plenty of such approaches (e.g.~those proposed in~\cite{DBLP:journals/tcad/AmyMMR13,MWZ:2011,matsumoto2008representation,WSOD:2013}) for different gate libraries like Clifford+T~\cite{boykin2000new} or the the NCV library~\cite{BBC+:95}. 
Those solutions can easily be integrated in compilers such as the one proposed here.

However, the second step represents a bigger challenge: Here, we need to determine a mapping of the $n$ logical qubits occurring in the quantum circuit (denoted by $q_0,q_1,\ldots,q_{n-1}$ in the following) to the $m\ge n$ physical qubits in the hardware (denoted by $Q_0,Q_1,\ldots,Q_{m-1}$ in the following) such that the physical (architectural) constraints reviewed above are satisfied. In almost all cases, it is not possible to determine such a mapping so that these constraints are satisfied for \emph{all} gates/operations throughout the circuit. Consequently, the mapping has to change dynamically. This can be achieved by adding SWAP gates to the circuit, which exchange the state of two physical qubits and, thus, allow to ``move'' the logical qubits to positions where they can interact with each other. 

\begin{figure}
	\begin{center}
		\footnotesize
		\mbox{
			\Qcircuit @C=.75em @R=0.5em {
			\push{\rule{4em}{0em}} & \lstick{Q_0 \shortleftarrow q_0} & \qswap      & \qw & \push{q_1} & \push{\rule{.3em}{0em}\phantom{\equiv}\rule{.8em}{0em}} & \ctrl{1} & \targ & \ctrl{1} & \qw & \push{\rule{.3em}{0em}\phantom{\equiv}\rule{.8em}{0em}} & \ctrl{1} & \gate{H} & \ctrl{1} & \gate{H} & \ctrl{1} & \qw\\
			\push{\rule{4em}{0em}} & \lstick{Q_1 \shortleftarrow q_1} & \qswap \qwx & \qw & \push{q_0} & \push{\rule[-5.5mm]{.3em}{0em}\equiv\rule{.8em}{0em}} & \targ & \ctrl{-1} & \targ & \qw & \push{\rule[-5.5mm]{.3em}{0em}\equiv\rule{.8em}{0em}} & \targ & \gate{H} & \targ & \gate{H} & \targ & \qw \\
		}}
	\end{center}	

		\vspace*{-3mm}
	\caption{Decomposition of a SWAP gate}
	\label{fig:swap}
		\vspace*{-3mm}
\end{figure}
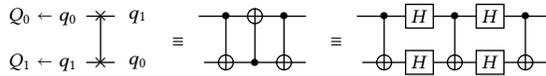

\begin{myexample}
Fig.~\ref{fig:swap} shows a SWAP operation and how it can be decomposed into operations that are available on IBM QX architectures. In the left-most circuit shown in Fig.~\ref{fig:swap}, the logical qubits $q_0$ and $q_1$ are mapped to the physical qubits $Q_0$ and $Q_1$, respectively. By applying a SWAP operation between $Q_0$ and $Q_1$ the ``position'' of $q_0$ and $q_1$ is permuted. The SWAP operation can be decomposed into a sequence of three CNOTs as shown in the center of Fig.~\ref{fig:swap}. If we assume that only CNOTs with control qubit $Q_0$ and target qubit $Q_1$ are possible (like for \emph{IBM QX2}; cf.~Fig.~\ref{fig:qx2arch}), we additionally have to invert the direction of the middle CNOT by applying Hadamard gates $H = U(\pi/2,0,\pi)$ before and after the CNOT (as shown in the right-most circuit in Fig.~\ref{fig:swap}).
\end{myexample}

Obviously, the number of additional SWAP operations shall be kept as small as possible, since each further operation decreases the fidelity of the circuit when running on an IBM QX device.\footnote{Note that these additional SWAPs also increase the depth of the circuit and, thus, its execution time on a quantum computer.} Therefore, IBM has set the goal to develop a compiler (including a mapping strategy) such that a circuit with the largest possible fidelity results~\cite{ibmDeveloperChallenge}. 

\section{State of the Art and Motivation for a Dedicated Solution}
\label{sec:motivation}

In this section, we discuss the current state of the art and motivate the need for a dedicated approach for compiling the circuits reviewed in Section~\ref{sec:circuits} to IBM's QX architectures reviewed in Section~\ref{sec:architectures}.

In the literature, there have already been several works that consider the mapping of quantum circuits to physical devices. However, most of them either focus on so-called \emph{nearest neighbor constraints} 
only~\cite{DBLP:journals/tcad/WilleLD14,DBLP:conf/aspdac/ShafaeiSP14,DBLP:conf/aspdac/WilleKWRCD16,zulehner2017exact,bhattacharjee2017depth} and/or on special quantum circuits to be mapped~\cite{fowler2004implementation,DBLP:journals/corr/VenturelliDRF17}. In the corresponding nearest neighbor architectures, a \mbox{2-qubit} gate can be applied to \emph{any} neighboring qubits and also in any desired direction---clearly violating the constraints for IBM QX architectures represented by the coupling maps. Moreover, many of the previously proposed approaches are only applicable for a very limited number of qubits (even lower than the 16 already available from IBM). 

In contrast, few methods exist which map the logical qubits of a quantum circuit to the physical ones of the IBM QX architectures. More precisely, 
a solution developed by IBM itself (based on Bravyi's algorithm and implemented in IBM's own SDK \emph{QISKit}~\cite{qiskit}) as well as the works presented in~\cite{zulehner2017efficient,siraichi2018qubit} is available thus far.
While the solution proposed in~\cite{siraichi2018qubit} has only been thoroughly evaluated for 5-qubit architectures and rather small circuits (and yields circuits with larger overhead than IBM's solution for 16-qubit devices), the approach proposed in~\cite{zulehner2017efficient} has shown significant improvements regarding gate count, depth, and runtime---clearly outperforming IBM's solution e.g.~on the 16-qubit architectures and for circuits composed of thousands of gates. 

This difference in quality is mainly because IBM's solution randomly searches  for a mapping that satisfies the physical constraints---leading to a rather small exploration of the search space so that only rather poor solutions are usually found. In contrast, the approach proposed in~\cite{zulehner2017efficient} aims for an optimized solution by exploring a larger part of the search space and additionally exploiting information of the circuit. More precisely, a look-ahead scheme is employed that considers gates that are applied in the near future and, thus, allows to determine mappings which constitute a local optima with respect to the number of SWAP operations. However, this solution is hardly suitable for the SU(4) circuits reviewed in Section~\ref{sec:circuits}, because: 

\begin{itemize}
\item The solution rests on the main idea to first divide the circuit into layers of gates\footnote{A layer contains only gates that act on disjoint qubits. Thus all gates of a layer can be applied in parallel.} and, afterwards, determine a permutation of qubits for each layer which satisfies all  physical (architectural) constraints within this subset of gates.\footnote{Between the respective layers, SWAP gates as shown in Fig.~\ref{fig:swap} are introduced to establish the respective qubit permutations.}

\item SU(4) circuits are composed of layers of gates which frequently contain $\frac{n}{2}$ different CNOT configurations (with~$n$ being the number of qubits). This is basically a worst case scenario since the more CNOT gates are employed within a layer, the more constraints have to be satisfied by a permutation of qubits.
\end{itemize}

As a consequence, the solution proposed in~\cite{zulehner2017efficient} cannot unfold its power for determining mapped circuits with smaller overhead than IBM's solution 
when applied for SU(4) circuits as it basically has to check all permutations within a layer until one is determined satisfying all constraints imposed by the CNOTs. Considering that SU(4) circuits have explicitly been provided by IBM to benchmark compilers, this is a serious drawback and motivates a compilation approach dedicated to this kind of circuits.

\section{Proposed Approach}
\label{sec:proposed}

In this section, we describe a dedicated procedure to compile SU(4) quantum circuits to IBM QX architectures. To overcome the limitations of the approach proposed in~\cite{zulehner2017efficient}, while keeping the availability of a look-ahead scheme, we break out of the \mbox{layered-based} approach and consider each gate on its own. In order to deal with the correspondingly resulting complexity, the proposed algorithm employs a combination of three steps: a pre-process step (reducing the complexity beforehand), a powerful search method (solving the mapping problem), and eventually a dedicated post-mapping optimization (exploiting further optimization potential after the mapping). 

\subsection{Pre-Process: Grouping Gates}
\label{sec:grouping}

Since each gate is considered on its own, the mapping may change after each gate (requiring much more calls of the mapping algorithm). To overcome this issue, we perform a pre-processing step where we form groups of gates, which we represent as a directed acyclic graph (DAG). By this, the mapping algorithm has to be called (at most) only once per group instead of once per gate. As further advantage, this DAG representation inherently encodes the precedence of the groups of gates and, thus, unveils important information about which groups of gates commute---giving the degree of freedom to choose which group shall be mapped next.

In order to group the gates, we topologically sort the circuit and group all gates that act on pairs of logical qubits (e.g.~on qubits $q_i$ and $q_j$) into a group $G_k$. This includes single qubit gates on $q_i$ or $q_j$ as well as CNOTs with control $q_i$ and target $q_j$ (or vice versa).
This grouping is done in a greedy fashion---until observing a CNOT with control or target $q_i$ ($q_j$) that acts on a qubit different from $q_j$ ($q_i$). This is possible, since gates that act on distinct sets of qubits are commutative.

\begin{myexample}
	Consider again the circuit shown at the right-hand side of Fig.~\ref{fig:kak}. Since, all gates of the circuit act on qubits $q_0$ and $q_1$, the grouped circuit contains a single group. By this, the mapping has to be changed at most once in order to apply all gates.
\end{myexample}

As stated above, grouping gates has a positive effect on the following mapping algorithm, since all gates of a group can be applied once the physical constraints are satisfied for the involved qubits.\footnote{Note that the direction of the CNOTs might have to be adjusted (which is rather cheap since only Hadamard gates have to be added).} Thus, the mapping of the gates of the circuit reduces to mapping the groups. 

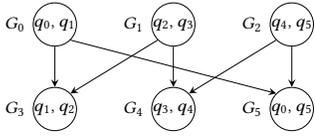
\begin{figure}
\centering

\scalebox{0.9}{\begin{tikzpicture}[>=stealth]
\matrix[matrix of nodes,ampersand replacement=\&,every node/.style={vertex},column sep={1.75cm,between origins},row sep={1.25cm,between origins}] (qmdd) {
	\node (n11) {$q_0$, $q_1$}; \& \node (n12) {$q_2$, $q_3$}; \& \node (n13) {$q_4$, $q_5$}; \\
	\node (n21) {$q_1$, $q_2$}; \& \node (n22) {$q_3$, $q_4$}; \& \node (n23) {$q_0$, $q_5$}; \\
};

\node[left=0cm of n11] {$G_0$};
\node[left=0cm of n12] {$G_1$};
\node[left=0cm of n13] {$G_2$};
\node[left=0cm of n21] {$G_3$};
\node[left=0cm of n22] {$G_4$};
\node[left=0cm of n23] {$G_5$};

\draw[->] (n11.south) -- (n21.north);
\draw[->] (n11.south east) -- (n23.north west);
\draw[->] (n12.south) -- (n22.north);
\draw[->] (n12.south west) -- (n21.north east);
\draw[->] (n13.south west) -- (n22.north east);
\draw[->] (n13.south) -- (n23.north);

\end{tikzpicture}}

		\vspace*{-3mm}
\caption{DAG after grouping the gates of the circuit}
\label{fig:dag}
		\vspace*{-3mm}
	
\end{figure}

\begin{myexample}
	Consider the DAG shown in Fig.~\ref{fig:dag}. This DAG represents a quantum circuit composed of 6 qubits, where the first layer is composed of  SU(4) gates between the logical qubits $q_0$ and $q_1$, $q_2$ and $q_3$, as well as $q_4$ and $q_5$, respectively. Moreover, the second layer contains SU(4) gates between the logical qubits $q_1$ and $q_2$, $q_3$ and $q_4$, as well as $q_0$ and $q_5$, respectively.
\end{myexample}

\subsection{Solving the Mapping Problem}
\label{sec:mapping}

After grouping the gates, the physical constraints of the target architecture given by the coupling map are satisfied by a mapping algorithm that determines a dynamically changing mapping of the logical qubits to the physical ones.
In theory, the mapping can change (by inserting SWAP gates) after each group---resulting in a huge search space since $m!$ possibilities exist for each such intermediate mapping. To cope with this enormous search space we use an A* search algorithm to find a solution that is as cheap as possible.  

For the mapping strategy presented in this paper, we choose an arbitrary initial mapping such that the physical constraints are satisfied for all groups in the DAG that do not have any predecessors (i.e.~the corresponding logical qubits are mapped to physical ones that are connected in the coupling map). By this, we can immediately add the gates of these groups to the (initially empty) compiled circuit.\footnote{Note that the qubits have to be relabeled according to the mapping and that the direction of some CNOTs might be adjusted.}

\begin{myexample}
Consider again the DAG in Fig.~\ref{fig:dag}, which describes the gate groups to be mapped. Assume that the circuit shall be compiled for the IBM~QX5 architecture, whose coupling map is depicted in Fig.~\ref{fig:qx5arch}. One possible initial mapping is $Q_1 \shortleftarrow q_0$, $Q_0 \shortleftarrow q_1$, $Q_2 \shortleftarrow q_4$, $Q_{15} \shortleftarrow q_2$, $Q_3 \shortleftarrow q_5$, and $Q_{14} \shortleftarrow q_3$ (i.e.~the logical qubits are mapped to the six left-most physical qubits). Using this initial mapping, the gate groups in the first layer (i.e.~$G_0$, $G_1$, and $G_2$) 
 can be applied since the involved logical qubits are mapped to physical ones that are connected in the coupling map for each of the groups.
\end{myexample}

After determining an initial mapping, the actual mapping procedure is composed of two alternating steps that are employed until all groups are mapped.

The first step adds all groups to the compiled circuit, whose parents in the DAG are already mapped and whose logical qubits are mapped to physical ones that are connected in the coupling map. 

\vspace*{-1mm}
\addtocounter{myexample}{-1}
\begin{myexample}[continued]
	The initial mapping additionally allows to add gates of group $G_3$  to the compiled circuit, since the its parents in the DAG (i.e.~the groups $G_1$ and $G_2$) are already mapped and the physical constraints are also satisfied (since $Q_0 \shortleftarrow q_1$ and $Q_{15} \shortleftarrow q_2$).
\end{myexample}
\vspace*{-1mm}

The second step determines the set of groups~$G_{next}$ that can be applied next according to their precedence in the circuit, i.e.~the set of groups whose parents in the DAG are already compiled. Then, the task of the mapping algorithm is to determine a new mapping (by inserting SWAP gates) such that the physical constraints are satisfied for at least one of the gate groups in $G_{next}$. 

\vspace*{-1mm}
\addtocounter{myexample}{-1}
\begin{myexample}[continued]
	One possibility is to incorporate a SWAP operation on the physical qubits $Q_{15}$ and $Q_{2}$ since this ``moves'' the logical qubits $q_3$ and $q_4$ towards each other and, thus, allows to add the gates from gate group $G_4$ to the compiled circuit. Finally, inserting another SWAP operation between the physical qubits $Q_{1}$ and $Q_{2}$ allows to add the gates of the group $G_5$ to the compiled circuit. Overall, two SWAP gates were inserted during the mapping procedure of the circuit.
		
	Another solution would be to incorporate a SWAP operation on the physical qubits $Q_2$ and $Q_3$. Since this ``moves'' the logical qubits $q_0$ and $q_5$, as well as the logical qubits $q_3$ and $q_4$ towards each other, the gate groups $G_4$ and $G_5$ can be applied by inserting a single SWAP operation during the compilation procedure. 
\end{myexample}
\vspace*{-1mm}

Among the solutions found by the mapping algorithm, we aim for determining the mapping that yields the lowest cost. Since there are $m!$ different mappings of the physical qubits, we utilize an A* search to avoid exploring the whole search space. The general idea of the A* search algorithm is to reach a \emph{goal state} from an \emph{initial state} such that the costs for reaching this state is the minimum (with respect to a certain heuristic). To this end, all successor states of the cheapest state are added to the explored search space (i.e.~the cheapest state is \emph{expanded}) until a \emph{goal state} is reached. The \emph{costs} $c(x) = f(x) + h(x)$ are thereby defined as the sum of the \emph{fix costs} $f(x)$ (i.e.~the costs for reaching the state $x$ from the initial state) and the \emph{heuristic costs} $h(x)$ (i.e.~an estimation for reaching a goal state from state~$x$).

This general description of the A* search algorithm has been  adjusted for the considered mapping problem. More precisely, the \emph{initial state} is the current mapping of the logical qubits to the physical ones. A \emph{goal state} is any state that describes a mapping where the physical constraints are satisfied for at least one of the groups groups.  \emph{Expanding} a state is conducted by  applying one SWAP operation between two physical qubits which results in a  successor mapping. Given that, the corresponding cost functions $f(x)$ and $h(x)$ have to be determined. The fix cost $f(x)$ of a state is given by the number of SWAP operations that have been added (starting from the current mapping). For the estimation of the remaining costs $h(x)$, the utilized heuristic employs a look-ahead scheme, which allows to significantly reduce the costs of the compiled circuit.

More precisely, for each group, we determine the distance of the physical qubits in the coupling map where the respective logical qubits are mapped to, and sum these distances up for all groups in $G_{next}$.\footnote{Note that the heuristic is not admissible and, hence, may not lead to a locally optimal solution. However, locally optima are not desired anyways, since these often yield to globally larger overhead.} 
By this, we do not only focus on one of these groups, but additionally try to optimize the mapping for groups that are applied in the near future.

\vspace*{-1mm}
\addtocounter{myexample}{-1}
\begin{myexample}[continued]
	The look-ahead scheme determines the goal node reached by conducting a SWAP operation between the physical qubits $Q_2$ and $Q_3$, since from the two solutions resulting in a goal state with costs 1 (inserting a single SWAP gate; as discussed above), the solution with the lower look-ahead costs was chosen.
\end{myexample}

\subsection{Post-Mapping Optimization}
\label{sec:optimization}

After satisfying the physical constraints given by the target architecture, we finally apply a dedicated \mbox{post-mapping} optimization in order to further reduce the costs of the compiled circuit. To this end, we regroup the gates of the compiled circuit as described in Section~\ref{sec:grouping}, since the mapping algorithm has added several SWAP gates to the compiled circuit. Then, we traverse the resulting DAG and optimize each group individually.

The key idea of the proposed optimization is that the functionality of the gates in a group $G_i$ can be represented by a single matrix from SU(4). Hence, we can easily build up this matrix by multiplying the unitary matrices representing the individual gates and, again, use \mbox{KAK-decomposition}~\cite{vatan2004optimal} to determine another group $G_i'$ with 3 CNOTs and 7 single qubit gates that realizes the same functionality (cf.~Section~\ref{sec:circuits}). If the gates in $G_i'$ have lower costs than the gates in the original group~$G_i$, we replace $G_i$ with $G_i'$ in the DAG. This especially works well, when applying a SWAP gate to two qubits, to which a gate from SU(4) has been applied right before.

\begin{myexample}
	Consider again the KAK-decomposition shown in Fig.~\ref{fig:kak} with its 3 CNOTs and 7 single qubit gates. Furthermore, assume that immediately afterwards a SWAP operation is applied to the physical qubits currently holding the logical qubits $q_0$ and $q_1$---yielding a group $G_i$ with 6 CNOTs and 11 single qubits.
	However, representing the overall functionality of this group as a unitary matrix from SU(4) and applying KAK-decomposition again yields another group~$G_i'$ with, again, 3 CNOTs and 7 single qubit gates. Hence, the SWAP operation can be conducted ``for free''. 
\end{myexample}

Note that the knowledge of this post-mapping optimization can be used to improve the mapping algorithm itself. More precisely, knowing that SWAP operations directly applied after a gate from SU(4) are ``for free'' can be included in the costs function $f(x)$ of the fix costs by setting the costs of the respective SWAP operation to 0.

Finally, a similar (but simpler) optimization can be applied for optimizing subsequent single qubit gates within a group. Such gates may e.g.~occur when swapping the direction of a CNOT by inserting Hadamard gates. Again, the $2\times 2$ unitary matrices describing the individual gates can be multiplied. Afterwards, the Euler angles of the rotations around the $z$ and $y$ axis are determined.

\begin{myexample}
	Consider again the KAK-decomposition shown in Fig.~\ref{fig:kak}. To change the direction of the center CNOT gate, Hadamard gates are inserted to each qubit before and after the CNOT---yielding to subsequent single qubit gates that are applied to $q_0$ and $q_1$, respectively. Again, this sequence of e.g.~$U_3$ and $H$ can again be represented by one single qubit gate.
\end{myexample}

\section{Experimental Evaluation}
\label{sec:exp}

\begin{figure*}
	\begin{subfigure}[b]{0.32\linewidth}
		\centering
		\includegraphics[width=\linewidth]{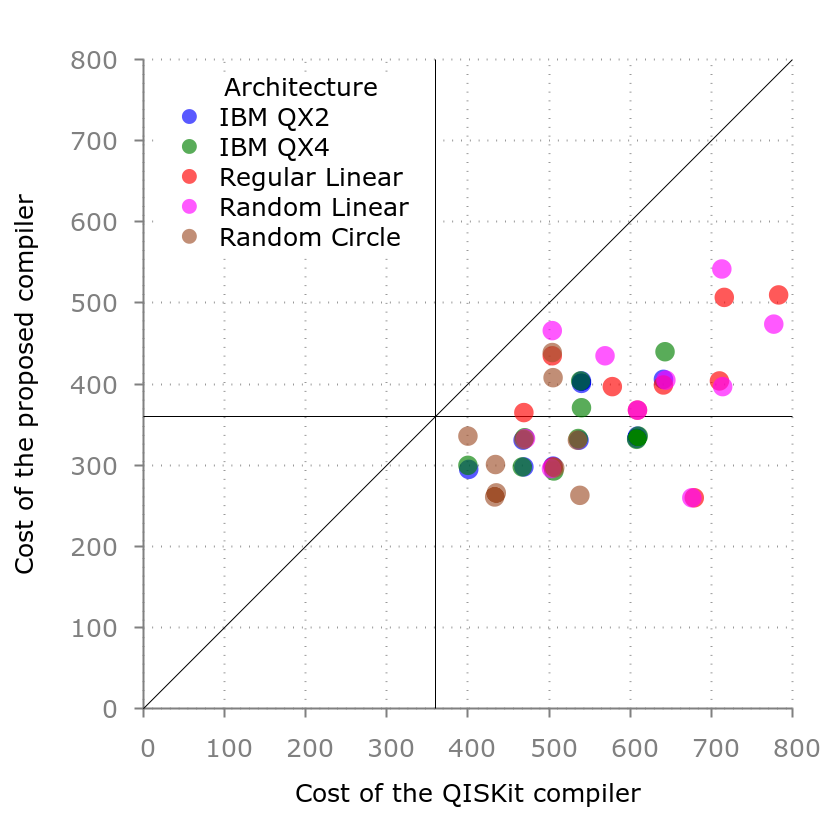}
	
		\vspace*{-3mm}
		\caption{5 qubit architectures}
		\label{fig:scatter_5q}
	\end{subfigure}\hfill
	\begin{subfigure}[b]{0.32\linewidth}
		\includegraphics[width=\linewidth]{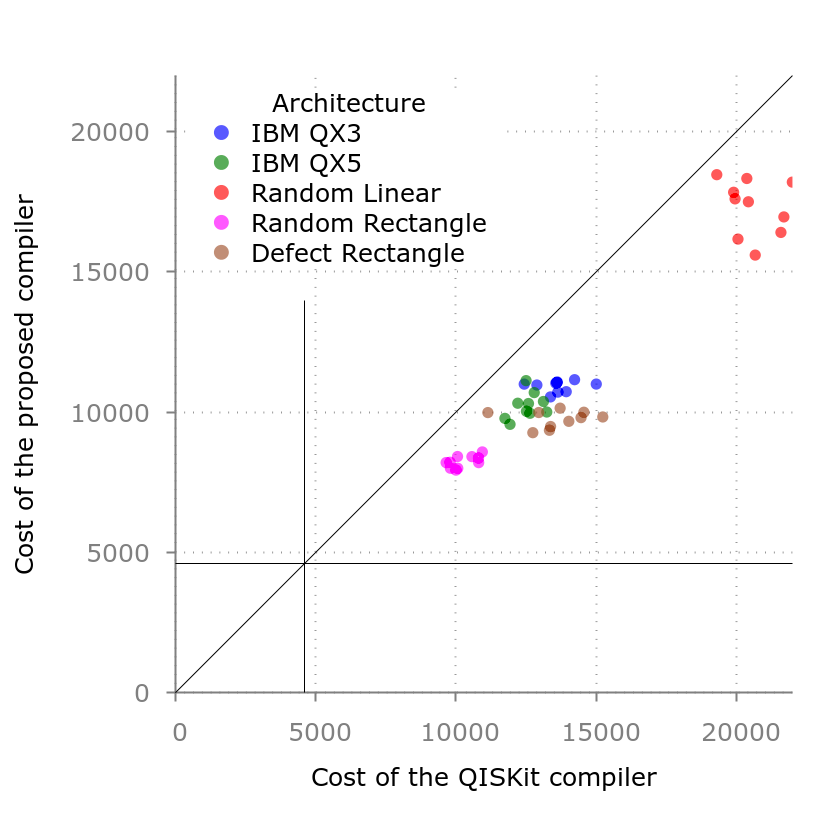}
	
		\vspace*{-3mm}
		\caption{16 qubit architectures}
		\label{fig:scatter_16q}
	\end{subfigure}\hfill
	\begin{subfigure}[b]{0.32\linewidth}
		\includegraphics[width=\linewidth]{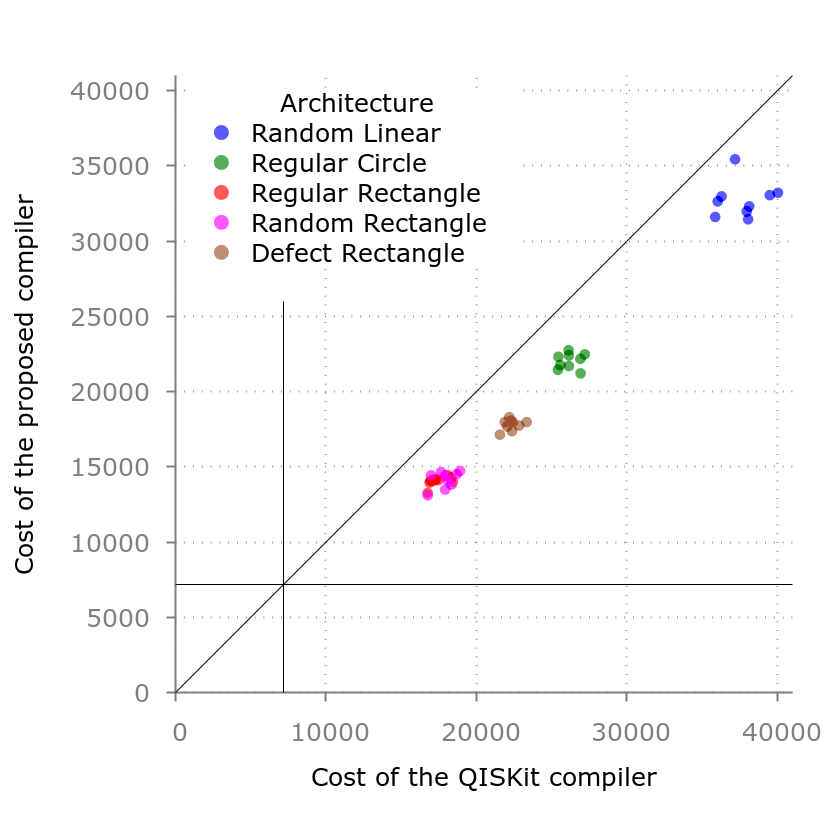}
	
		\vspace*{-3mm}
		\caption{20 qubit architectures}
		\label{fig:scatter_20q}
	\end{subfigure}

		\vspace*{-3mm}
	\caption{Cost of the compiled circuits}
	\label{fig:scatter}
		\vspace*{-3mm}
\end{figure*}

In this section, we experimentally evaluate the proposed approach and compare it to the compiler available in IBM's SDK \emph{QISKit}~\cite{qiskit}.\footnote{Note that no experimental comparison is reported for the approach presented in \cite{zulehner2017efficient} since, as discussed in Section~\ref{sec:motivation}, the SU(4) circuits represent a worst case for them and, hence, this method is not feasible for those benchmarks. In fact, running the publicly available implementation (taken from \mbox{\url{http://iic.jku.at/eda/research/ibm_qx_mapping/}}) confirms that this method frequently times out for those benchmarks. For the same reason, also no results for the approach presented in~\cite{siraichi2018qubit} is presented which is applicable for a rather tiny number of qubits only, respectively.}
To this end, we implemented the proposed methodology in Cython (available at~\url{http://iic.jku.at/eda/research/ibm_qx_mapping}) and used the scripts provided by IBM to conduct the evaluation (these scripts are available at~\cite{ibmDeveloperChallenge}). Since the fidelity of CNOT gates is approximately 10 times lower for IBM QX architectures than the fidelity of single qubit gates (cf.~\cite{ibmDevices}), the provided cost function assigns a cost of 10 for each CNOT as well as a cost of 1 for single qubit gate.\footnote{Note that a single qubit gate $U(0,0,\lambda)$ has cost 0 since no pulse is applied to the respective qubit in this case.} All evaluation have been conducted on a 3.8\,GHz machine with 32\,GB~RAM.

Besides the circuits, IBM also provides several coupling maps for architectures with 5, 16, and 20 qubits, respectively. These architectures include the existing quantum devices \emph{IBM QX2}, \emph{IBM~QX4}, and \emph{IBM QX5}, as well as other architectures where the qubits are arranged in a linear, circular, or rectangular fashion. 
For these architectures, the direction of the arrows in the coupling maps are chosen randomly by IBM (or connections are missing at all) to provide a realistic basis for the evaluation. 
For each number of qubits in the architectures (5, 16, or 20), we use 10 circuits, which we compile to each architecture with the corresponding number of qubits. Eventually, this results in a setting which is also used by IBM to evaluate compilers submitted to the QISKit Developer Challenge~\cite{ibmDeveloperChallenge}. The resulting costs are visualized by means of scatter plots in Fig.~\ref{fig:scatter}.

Each of the plots in Fig.~\ref{fig:scatter} shows the cost of the compiled circuits when using the QISKit compiler on the \mbox{x-axis}, as well as the cost of the compiled circuit when using the proposed solution on the y-axis. Each point represents one SU(4) circuit that is compiled for a certain architecture. Hence, a point underneath the main diagonal, indicates the proposed solution yields a circuit with lower cost (which is the case for all evaluated circuits and architectures). The larger the distance to the main diagonal, the larger the improvement. We additionally added horizontal and vertical lines that indicate the cost of the original circuits (i.e.~the cost before compilation).

As can be seen in Fig.~\ref{fig:scatter_5q}, circuits compiled by the proposed methodology may be cheaper than the original circuit (despite the fact that SWAP gates are \emph{added} during the compilation process). This is possible since, in some cases, two $SU(4)$ gates are subsequently applied to the same two qubits. By using our post-mapping optimization (cf.~Section~\ref{sec:optimization}), these gates can be combined to a single gate from $SU(4)$. Overall, we achieve an average improvement by a factor of 1.54 compared to IBM's own solution for the 5-qubit architectures. For the 16 and 20 qubit architectures, the probability that two subsequent $SU(4)$ gates are applied to the same qubits is almost zero. But although this does not allow as much post-mapping optimization as for the 5-qubit architectures, we still observe significant improvements of a factor of 1.26 and 1.22 on average, respectively. The precise improvements for each architecture are listed in Table~\ref{tab:results}. 

Besides the average improvement in terms of the provided cost function, the proposed method is also significantly faster than IBM's solution. While IBM's solution requires more than 200 seconds for mapping some of the circuits composed of 20 qubits, the proposed method was able to map each of the circuits within 10 seconds. On average, we obtain
an improvement of the runtime by a factor of 5.68, 16.42, and 21.90 for the architectures with 5, 16, and 20 qubits, respectively (cf.~Table~\ref{tab:results}).

Overall, the evaluation using the scrips, circuits, and coupling maps provided by IBM shows that the dedicated compile methodology proposed in this paper significantly outperforms IBM's own solution regarding the provided cost function (which estimates fidelity) as well as runtime. Moreover, the solution proposed in this paper has been declared winner of the QISKit Developer Challenge. According to IBM, it yields compiled circuits with at least 10\% better costs than the other submissions while generating them at least 6 times faster. 

\begin{table}
	\caption{Average improvement factors}
	\label{tab:results}
			\vspace*{-2mm}	
	\scriptsize
	\centering
	\setlength{\tabcolsep}{2pt}
	\begin{tabular}{l|c|c||l|c|c||l|c|c}
		\multicolumn{3}{c||}{5 qubits} & \multicolumn{3}{c||}{16 qubits} & \multicolumn{3}{c}{20 qubits} \\
		architecture & cost & time & architecture & cost & time & architecture & cost & time \\\hline\hline
		IBM QX2 & 1.55 & 5.96 & IBM QX3 & 1.25 & 14.85 & Random Linear & 1.15 & 14.64 \\
		IBM QX4 & 1.54 & 5.84 &	IBM QX5 & 1.23 & 12.87 & Regular Circle & 1.19 & 14.25 \\
		Regular Linear & 1.58 & 5.40 & Random Linar & 1.19 & 11.52 & Regular Rectangle & 1.24 & 32.67 \\
		Random Linear & 1.57 & 5.43 & Random Rectangle & 1.24 & 20.99 & Random Rectangle & 1.27 & 33.95 \\
		Random Circle & 1.49 & 5.81 & Defect Rectangle & 1.39 & 25.80 & Defect Rectangle & 1.25 & 21.78 \\\hline
		avg & 1.54 & 5.68 & avg  & 1.26 &16.42 & avg & 1.22 & 21.90 \\
	\end{tabular}
		\vspace*{-1mm}

\end{table}

\section{Conclusions}
\label{sec:conclusions}

In this paper, we presented a dedicated method for compiling circuits composed of $SU(4)$ gates to IBM QX architectures. By using a preprocessing-step that groups the gates in order to reduce the complexity, a mapping algorithm based on an A* search with a look-ahead scheme, as well as a dedicated post-mapping optimization, we were able to overcome the shortcomings of previously proposed approaches. Our evaluation using tools provided by IBM clearly shows that the proposed approach significantly outperforms the compiler available in IBM's SDK \emph{QISKit} regarding a cost function that estimates the fidelity of the compiled circuit as well as runtime. Moreover, it has been declared winner of the QISKit Developer Challenge. An implementation is publicly available at \url{http://iic.jku.at/eda/research/ibm_qx_mapping}.

\begin{acks}
	This work has partially been supported by the European Union through the
	COST Action IC1405.
\end{acks}

\bibliographystyle{abbrv}
{\bibliography{literature}}

\end{document}